%% file: draft_v2.tex
\documentclass[
superscriptaddress,
reprint,
 amsmath,amssymb,
 aps,
prl
]{revtex4-2}
\usepackage{booktabs, makecell}
\usepackage{braket}
\usepackage{graphicx}
\usepackage{dcolumn}
\usepackage{bm}
\usepackage{multirow}
\usepackage{amsmath,array,multirow,makecell,xcolor,colortbl}
\usepackage{hyperref}

\usepackage{color}
\usepackage{xcolor}

\newcommand{\bo}[1]{\mathbf{#1}}
\newcommand{\q}{\mathbf{q}}
\newcommand{\kk}{\mathbf{k}}
\newcommand{\G}{\mathbf{G}}
\newcommand{\PP}{\mathbf{P}}
\newcommand{\rr}{\mathbf{r}}
\newcommand{\uu}{\mathbf{u}}
\newcommand{\eh}{\hat{\mathbf{e}}}

\newcommand{\Rm}{\mathcal{R}}            
\newcommand{\rt}{\mathsf{t}}             
\newcommand{\Op}{\hat{\mathcal{O}}}      
\newcommand{\Xc}{\mathcal{X}}            
\newcommand{\Yc}{\mathcal{Y}}            

\newcommand{\supplementarysection}{%
  \setcounter{figure}{0}
  \let\oldthefigure\thefigure
  \renewcommand{\thefigure}{S\oldthefigure}
  \setcounter{section}{0}
  \let\oldthesection\thesection
  \renewcommand{\thesection}{S\oldthesection}
  \setcounter{equation}{0}
  \let\oldtheequation\theequation
  \renewcommand{\theequation}{S\oldtheequation}
  \setcounter{table}{0}
  \let\oldthetable\thetable
  \renewcommand{\thetable}{S\oldthetable}
}
\begin{document}
\title{Goldstone-Mediated Chiral Superconductivity in Self-Doped Electron Crystals}

\author{Shuchen Zhang}
\affiliation{Department of Physics, Carnegie Mellon University, Pittsburgh, Pennsylvania, 15213, USA
}
\author{Zhihuan Dong}
\affiliation{Department of Physics, Harvard University, Cambridge, Massachusetts, 02138, USA
}
\author{Zhaoyu Han}
\affiliation{Department of Physics, Harvard University, Cambridge, Massachusetts, 02138, USA
}

\author{Shubhayu Chatterjee}
\affiliation{Department of Physics, Carnegie Mellon University, Pittsburgh, Pennsylvania, 15213, USA
}
\date{\today}

\begin{abstract}
Motivated by the observation of superconductivity in rhombohedral graphene in the proximity of electronic crystalline phases, we study the possibility of pairing mediated by electronic phonons. 
We focus on the vicinity of the phase boundary between trivial and topological electron crystals, where self-doping generically gives rise to a small Fermi sea of carriers with Dirac dispersion. 
We compute the Goldstone phonon spectra of the self-doped crystal (SDC) using a self-consistent approach, and subsequently derive the effective electronic-phonon-mediated interaction between itinerant carriers at the Fermi surface.
Remarkably, we find that the Coulomb repulsion can be overscreened by the Goldstone phonons, leading to a chiral topological superconductor with an unpaired Majorana edge mode. 
Our conclusions are robust across the $\lambda$-jellium model and a simplified model of rhombohedral graphene, and we propose characteristic non-linear I-V signals as a test of our prediction.  
More generally, our results outline an intriguing route for unconventional superconductivity where both electronic correlations and topology play key roles.
\end{abstract}

\maketitle

The intricate interplay of electronic band topology and strong correlations in two dimensions can lead to novel quantum phases.
Examples of such phases include unconventional all-electronic superconductivity even when the inter-electronic interaction is purely repulsive~\cite{divic2025anyon,sahay2024superconductivity,kuhlenkamp2025robust,ShiSenthilPRX,shi2025non,CBZ2020,khalaf2021charged,CIZ2022,KwanPRX,pichler2026microscopic,guerci2024topological,herzog2026many,tavakol2026pairing,gallegos2026repulsion,gonccalves2026chiral}, as well as topological analogs of Wigner crystals~\cite{Wigner1934,bonsall1977some,Joy2025,skinner2026lecture} -- anomalous Hall crystals with a quantized Hall conductance~\cite{DongAHC2024,soejima2024anomalous,dong2024theory,Zhou2024,dong2024stability,tan2024parent,sheng2024quantum,zhou2025new,desrochers2026elastic}.
Recent experimental observations suggest a new phenomenon 
at the  boundary between a topologically trivial Wigner crystal and an anomalous Hall crystal: self doping~\cite{Feng2026SDC,Dong2026MWC}.
Specifically, in proximity to a charge-gap closing transition between two insulating electronic crystals, the system spontaneously develops a small density of itinerant carriers while retaining crystalline order with a slightly different unit cell size. 
This new phase, which features a small Fermi sea of carriers with Dirac dispersion moving on the background of a weakly incommensurate Wigner crystal, has been dubbed a self-doped crystal~\cite{Feng2026SDC} or a metallic Wigner crystal~\cite{Dong2026MWC}. 
This leads to a natural question: can collective vibrations of the electronic crystal lead to superconductivity of the itinerant electrons?
If so, how does the Dirac dispersion of the carriers and the topology of the parent crystalline phases affect the strength and symmetry of pairing?

Recent observations in rhombohedral tetralayer and pentalayer graphene provide experimental motivation to study electronic lattice vibration mediated pairing~\cite{Han2025Chiral}. 
In such systems, multiple superconducting regimes have been identified adjacent to insulating Wigner crystals. 
Most notably, the `SC3' regime in Ref.~\cite{Han2025Chiral} shows quantum oscillation signals indicating a carrier density that is about a tenth of the total carrier number with an opposite sign~\cite{han2026evidencemetallicwignercrystal}. 
This strongly supports an interpretation of superconductivity emerging from a hole-doped Wigner crystal. 
The `SC1' and `SC2' phases likewise lie inside regimes with anomalous quantum oscillation signals~\cite{kalantre2026fermiologycandidatechiralsuperconductor}, suggesting the presence of crystalline order in the parent state. 
Whether vibrational Goldstone modes of self-doped crystals can mediate chiral superconductivity, like other collective modes~\cite{dong2023superconductivity,dong2026superconductivity,Christos2026,Ghazaryan2021,CWBZ2022}, remains an open question. 

In this paper, we investigate the intriguing possibility of collective vibrations of electronic crystals as pairing mediators between mobile carriers in a self-doped crystal (SDC). 
We find that such \textit{electronic phonons} can overscreen the bare Coulomb repulsion and mediate pairing of carriers into a chiral topological superconductor within a self-consistent formalism. 
Remarkably, the same Coulomb repulsion that leads to Wigner crystallization effectively turns against itself via emergent crystal dynamics to give rise to pairing. 
The underlying chiral quantum geometry of the parent crystal plays a crucial role, enhancing attractive interactions in the p-wave channel where the bare Coulomb repulsion is weaker. 
As a result, chiral superconductivity occurs robustly in SDCs, both in the phenomenological $\lambda$-jellium model~\cite{Soejima2025} -- a continuum model with broken time-reversal symmetry and tunable Berry curvature that hosts an SDC, and a model Hamiltonian for rhombohedral multilayer graphene (RMG)~\cite{Park2023,dong2024theory,soejima2024anomalous,DongAHC2024,Zhou2024,han2025exactmodelschiralflatband}. 

Our key results are threefold. 
First, we extract the dispersion of the vibrational modes of SDCs using a Bethe-Saltpeter approach equivalent to time-dependent self-consistent Hartree-Fock~\cite{Vasiliou2024Goldstone}.
We integrate out the lowest branches of these phonon modes to obtain an effective electron-electron interaction following Ref.~\cite{Vasiliou2024Goldstone}, and show that the interaction can reach a non-negligible fraction of the Fermi energy $E_F$, i.e., McMillan's dimensionless measure of electron-phonon coupling strength on the Fermi surface~\cite{PhysRev.167.331} can be as large as $\sim 0.2$ in the $p$-wave channel. 
Next, we use this interaction to solve the Eliashberg equation for the superconducting gap function~\cite{Bultinck2024Nodal}, and find chiral pairing. 
Finally, we show the Bogoliubov de Gennes (BdG) Chern number $C_{\rm BdG}$ of the superconductor is odd in the regime of interest, indicating the presence of chiral Majorana edge modes~\cite{Schnyder_2009,sato2017topological,qi2011topological}. 
Our results provide evidence for a new route toward all-electronic topological superconductivity, and identify self-doped electronic crystals as a promising platform for realizing chiral superconductors, with potential relevance to rhombohedral multilayer graphene.

\paragraph{Pairing: how and why? --}
To get an intuitive understanding of the physics without explicit reference to specific models, it is useful to ask why a superconducting ground state can at all be stable in spite of purely repulsive interactions in an SDC, and what possible superconducting phases one may expect in such a setting.
We tackle the latter question first. 
Our consideration is based on the observation that an SDC appears as an intermediate phase between two electronic crystals of different Chern numbers, that are otherwise connected by a continuous Dirac transition~\cite{Feng2026SDC}.
Consider some microscopic parameter $g$ that tunes such
a Dirac transition at $g_D$ between two crystals with Chern
numbers $C+1$ and $C$.
In the presence of superconducting pairing, the $C+1$ ($C$) chiral fermionic edge states of a crystal with Chern number $C+1$ ($C$) evolve into $C_{\rm BdG} = 2C + 2$ ($2C$) chiral Majorana modes. 
However, the pairing also lifts the degeneracy between the two additional chiral Majorana edge modes, implying that there are two bulk gap closures (one when each mode moves into the bulk) and an intermediate topological phase with $C_{\rm BdG} = 2C + 1$.
This is precisely what we observe in our calculations across different models --- a fully gapped superconducting phase with $C_{\rm BdG} = 1$ between a topological ($C = 1$) and trivial ($C = 0$) electronic crystal.

\begin{figure}[ht]
\includegraphics[width=0.5\textwidth]{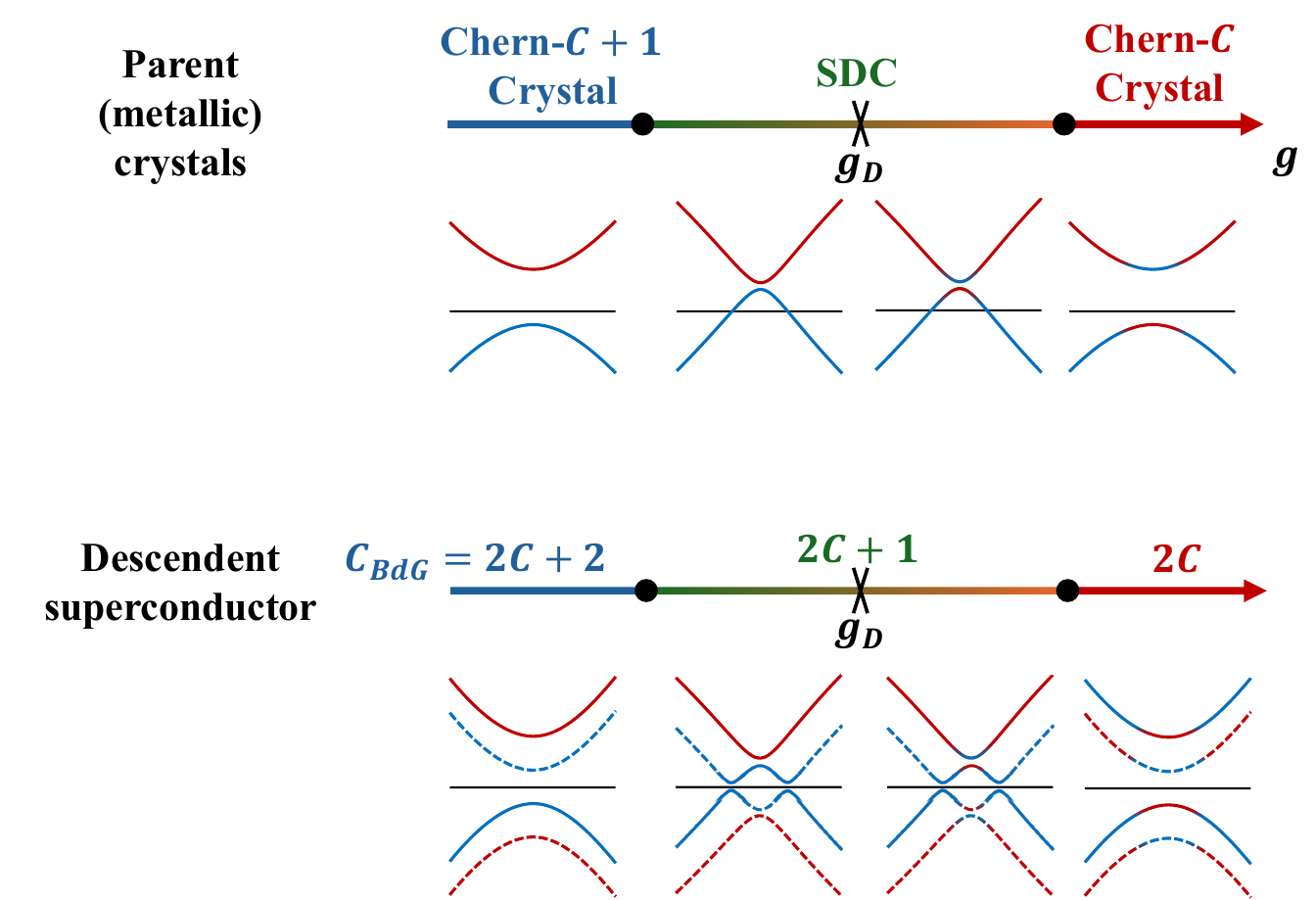}
\caption{Schematic phase diagrams: parent phases and descendant superconductors. Phase transitions are marked by black dots. The cross sign denotes the point of Dirac crossing in the parent crystal band, which is not associated with a physical phase transition.}
\label{sch_pd}
\end{figure}

Intuitively, the intermediate Majorana edge mode arises from a chiral winding $W$ of the order parameter, i.e., $C_{\rm BdG} = 2C + W$\footnote{Note that the winding number is gauge-dependent. Throughout this work, we will define winding number under the gauge choice that is single-valued within the dopant pocket.}. 
This may be understood from a momentum space Little-Parks effect~\cite{Dong_2024,MayMann2026Mechanism}, which suggests that winding number of pairing order is related to the Berry curvature $\Phi$ enclosed by the Fermi surface $W = [2 \Phi/2 \pi]$.
Note that the winding jumps at the Dirac transition point for the parent crystal band, but this is not a physical phase transition. 
The step jump of 2 in the winding number is purely a consequence of the gauge choice, which is single-valued within the dopant pocket. 
Its contribution to $C_{BdG}$ is canceled by the change of Chern number. 
This is immediately transparent if we think about the evolution of BdG band structure [Fig.~\ref{sch_pd}]. 
The Dirac band inversion only affects higher energy part of the Bogoliubov spectrum and does not close the quasiparticle band gap, leaving $C_{\rm BdG}$ unchanged. 
This physics is reminiscent of earlier proposals of topological superconductivity induced by proximity effect of an
s-wave superconductor on a quantum Hall state~\cite{QiHughesZhang} or the Dirac surface state of a three-dimensional topological insulator~\cite{FuKane2008}.

\begin{figure*}[!htbp]
\includegraphics[width=0.95\textwidth]{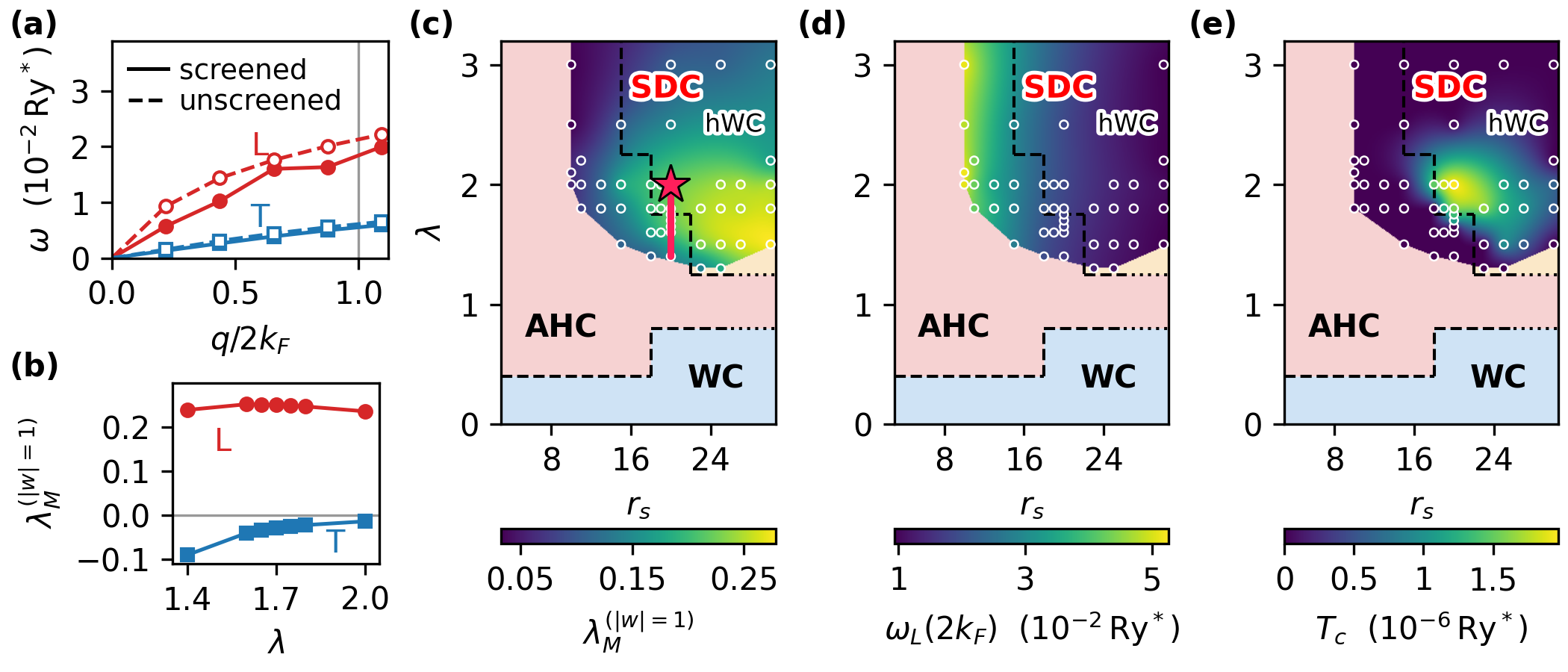}
\caption{(a) Dispersion of the Goldstone phonon modes of the $\lambda-$jellium model in the SDC at $(\lambda,r_s) = (2,20)$, marked by a red star in (c). Dotted lines show the corresponding dispersions in the parent electronic crystal. Screening causes the longitudinal mode (L) to become approximately linear at small momentum, while the transverse mode (T) is minimally affected. (b) The McMillan parameters of the longitudinal and transverse electronic phonon modes respectively, along a line cut in the $(\lambda, r_s)$ plane in the SDC, marked by a red line in (c). (c) The McMillan parameter $\lambda_M$ for winding number W = $\pm 1$ in the self-doped regime.
(d) The longitudinal phonon energy at $q = 2k_F$ in the self-doped regime.
(e) Superconducting critical temperature $T_c$ from solving Eliashberg equations. For (c)-(e), the color maps are interpolated from computations at the white circles, and the dashed lines correspond to phase boundaries between the parent crystals~\cite{Feng2026SDC}.The color maps roughly cover the self-doping regime in our numerics. } 
\label{fig:results}
\end{figure*}

Next, we turn to the question of why an SDC can potentially exhibit a superconducting ground state via coupling between itinerant carriers and electronic lattice vibrations. 
Intuitively, we expect density modulations of the electronic lattice arising from longitudinal vibrational modes to couple strongly to the itinerant carriers. 
At low momenta $q$, such longitudinal electronic phonons closely resemble two-dimensional plasmon modes, with energy $\omega_L(q) \propto \sqrt{q}$~\cite{Dong2025Phonons}.
Accordingly the group velocity $v_L = \partial_q \omega_L(q) \propto 1/\sqrt{q}$ increases as $q$ decreases, potentially allowing 
vibrations of the electronic crystal itself to dynamically overscreen the Coulomb repulsion between itinerant carriers with smaller Fermi velocity $v_F$. 
Indeed, a closely related mechanism has been proposed to mediate superconducting pairing in low-density ferroelectrics via critical optical phonon modes~\cite{KoziiPRX}. 

However, the \textit{electron-phonon} coupling in SDCs is intrinsically different from that in conventional solid state systems, due to the fact that the crystal in SDC are formed by electrons indistinguishable from the electrons forming the Fermi surface. 
Hence, a more nuanced treatment is needed, and our self-consistent estimate shows that the relevant longitudinal phonon velocity $v_L(q \approx 2 k_F)$ that scatters carriers across the small Fermi surface is typically an order of magnitude smaller than the Fermi velocity $v_F$. 
Therefore, crystal distortions relax on a slower scale, and carriers may attract each other by moving fast towards (or away from, depending on their sign) the crystal distortion before it can relax, analogous to the retarded phonon-mediated attraction in conventional superconductors~\cite{BCS}. 
What is somewhat surprising here is that the long-wavelength {\it acoustic} phonons couple strongly enough to mediate pairing on a small Fermi surface. 
We now illustrate this physics by explicitly considering the $\lambda$-jellium model~\cite{Soejima2025}, which features the SDC as the ground state~\cite{Dong2026MWC,Feng2026SDC}. 

\paragraph{$\lambda$-jellium model--} The $\lambda$-jellium model is a spinless, two-orbital continuum model designed so that the lower band has a tunable Berry curvature at \emph{fixed} dispersion~\cite{Soejima2025}. 
The Hamiltonian of the model is $H = H_0 + H_{\rm int}$, where the non-interacting Hamiltonian $H_0$ is given by
\begin{equation}
  H_0(\mathbf q) \;=\; \frac{\q^2}{2m^*}\, \mathbb{I}_2
  \;+\; \Delta\, M(\mathbf q),
  \qquad
  M(\mathbf q) = \begin{pmatrix} \lambda^2 q\bar q & -\lambda \bar q \\[2pt]
 -\lambda q       & 1 \end{pmatrix} ,
  \label{eq:B2}
\end{equation}
where $q = q_x + i q_y$ is a complex representation of the momentum $\q$, $m^*$ is the effective electron mass, and the band-gap set by $\Delta$ is taken to be large to allow us to focus only on the lower quadratic band. 
$H_0$ explicitly breaks spin-less time-reversal because of terms linear in momentum, and thus essentially behaves like the Hamiltonian of a \textit{quarter metal}. i.e., a spin and valley-polarized metal~\cite{zhou2021half,yang2025impact,hua2026multi,ding2026spin,WVZC2024}. 
For the interacting Hamiltonian $H_{\rm int}$, we use a dual gate-screened Coulomb interaction
\begin{equation}
V(\mathbf{q})=\frac{e^2}{2\epsilon}\frac{\tanh(|\mathbf{q}|d_g)}{|\mathbf{q}|}, 
\end{equation}
where $d_g$ is the gate-distance, and $\epsilon$ is the dielectric constant. 
This model has two tunable parameters: $\lambda$, which changes the distribution of Berry curvature $\Omega(\q) = 2\lambda^2/(1 + \lambda^2 |\q|^2)$ in the lower band without affecting its dispersion, and $r_s = a_\rho/a_0$, which tunes the ratio of the kinetic energy $\hbar^2/m^* a_\rho^2$ to the potential energy $e^2/\epsilon a_\rho$ (where $\rho = \pi a_\rho^2$ sets the electron density $\rho$, and $a_0 = \epsilon \hbar^2/m^* e^2$ is the effective Bohr radius). 
As a function of $(\lambda, r_s)$, self-consistent Hartree-Fock numerics finds an SDC near the phase boundary between an anomalous Hall crystal and a halo-Wigner crystal~\cite{Feng2026SDC,Dong2026MWC}.

\paragraph{Vibration mediated electronic interactions-}
We now consider the region of the phase diagram in the $(\lambda, r_s)$ plane which exhibits an SDC with Dirac carriers near the $\Gamma$ point ($\q = 0$).
To obtain the collective mode spectra, we follow the formalism of Ref.~\cite{Vasiliou2024Goldstone}, and solve a generalized eigenvalue problem based on a zero-temperature Bethe-Saltpeter equation (BSE)~\cite{SM}. 
The general solution to the BSE at each momentum $\q$ includes Goldstone modes (electronic phonons) that correspond to two spontaneously broken translation symmetries, as well as other collective modes such as particle-hole excitations at the Fermi surface. 
To isolate the Goldstone modes at each $\q$, we therefore choose the BSE eigenmode that has the strongest overlap with a physical displacement of the crystal at that wave-vector. 
Relegating a more precise formulation of this criteria to the Supplementary Material (SM~\cite{SM}), we find that we can isolate the expected longitudinal and transverse phonon modes at low momenta.
We plot their spectra in Fig \ref{fig:results}(a).
As expected, we find that the transverse phonon is very weakly affected by screening due to self-doped carriers and inter-band particle-hole excitations, but the longitudinal mode becomes approximately linear at low momenta ($q a_\rho \ll 1$)  as a consequence of screening. 

Next, we proceed to compute the scattering vertex between the itinerant electrons and the Goldstone modes following the methodology in Ref.~\cite{Vasiliou2024Goldstone},  and subsequently integrate them out to arrive at an effective frequency($\nu$)-dependent electron-electron interaction within the low energy subspace consisting of the lowest two electronic bands. As an indicator of the strength of the phonon-mediated interaction on the Fermi surface~\cite{PhysRev.167.331}, we plot McMillan's dimensionless parameter, which we call $\lambda_M$, in the leading angular momentum channel (see SM for definition~\cite{SM}) in Fig.~\ref{fig:results}(c). 
The mode-resolved version of this quantity is plotted in Fig.~\ref{fig:results}(b) along a line-cut where superconductivity is seen [Fig.~\ref{fig:results}(e)], clearly demonstrating that the longitudinal mode plays a dominant role in mediating attraction.
Further, since the relevant phonon velocity $v_L (q \approx 2 k_F)$ is typically an order of magnitude smaller than $v_F$ in our self-consistent solution, we need to account for retardation effects in our study of superconductivity, that we turn to next.



\paragraph{Superconductivity--} 
We solve the Eliashberg equation for the superconducting gap function~\cite{Marsiglio_2020} by linearizing it at the Fermi surface, and projecting the interaction to the lowest Hartree-Fock band~\cite{SM}. 
This is formally justified by a small Migdal parameter in the entire investigated parameter regime~\cite{SM}. 
Spin is polarized in our model, so the solution must be odd-parity due to the fermion statistics.
Since time-reversal is broken by the parent state, one may expect a chiral pair wave-function. 
Indeed, we find a fully gapped chiral superconductor, with a highest critical temperature of $T_c \approx 2 \times 10^{-6}$ $Ry^*$, where $Ry^*$ = $m^* e^4/(2\hbar^2 \epsilon^2)$. 
Choosing typical values $m^* = 5 m_e$ and $ \epsilon = 4$ leads to $T_c \approx 94$ mK.

The full landscape of $T_c(\lambda, r_s)$ is shown in Fig.~\ref{fig:results}(e). 
The behavior of $T_c$ qualitatively agrees with the expectations from Eliashberg theory: $T_c$ scales favorably both with increasing the effective electron-phonon coupling strength $\lambda_M$ and the relevant phonon frequency, which in our case is the longitudinal phonon frequency $\omega_L(2k_F)$, as plotted in Fig.~\ref{fig:results}(d). 
The highest $T_c$ appears to be reached by
a balance between these two quantities, rather than being determined by one of them alone.


\paragraph{Pairing angular  momentum and BdG Chern number-} 
We now turn to a discussion of the properties of the superconductor.
For generality, we consider scenarios with $C_n$ rotation symmetries ($n=6$ for $\lambda$-jellium and $n=3$ for RMG). 
In both cases, we find a small hole pocket at $\gamma$ point of the crystal Brillouin Zone (BZ). 
We label the pairing channels with angular momentum of Cooper pair,
\begin{align}
    L_{pair} \equiv 2L_\gamma + W \mathrm{(mod\;n)},
\end{align}
where $L_\gamma$ is the rotation symmetry index of parent electron crystal band at $\gamma$ point, and $W$ is the phase winding of the pairing order parameter $\Delta(\mathbf{k})$ around the Fermi surface, extracted under the gauge convention that the Bloch function stays single-valued inside the hole pocket. 
Notably, with this gauge convention for $W$, $L_{pair}$ stays a constant as the parent crystal band goes through a Dirac transition, where the symmetry index at $\gamma$ jumps. 
The single-valued nature of our gauge choice demands the winding $W$ to adjust in response to the change of $L_\gamma$. 
Then, $C_{BdG}$ is related to the pairing angular momentum through
\begin{align}
    &\mathrm{With\;} C_6:\; C_{BdG} = 2C+W \equiv 2 (2L_\kappa + 3L_{\cal M}) + L_{pair} \mathrm{(mod\;6)}\nonumber \\
    &\mathrm{With\;} C_3:\; C_{BdG} = 2C+W \equiv 2 (L_\kappa + L_{\kappa'}) + L_{pair} \mathrm{(mod\;3)}
\end{align}
With this convention, the chiral superconductor descending from the SDC in $\lambda$-jellium shows $L_{pair}=1$ and $C_{BdG}=1$, consistent with the expectation outlined in Fig.~\ref{sch_pd}.


\begin{figure}[!t]
\includegraphics[width=0.5\textwidth]{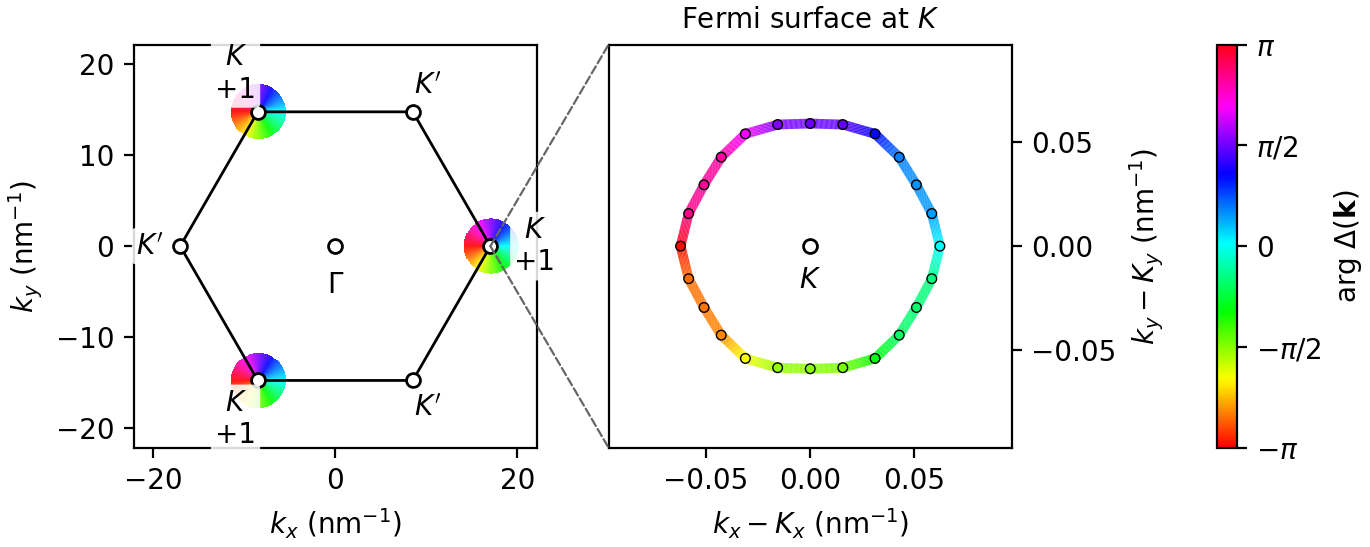}
\caption{Schematic of chiral pairing in small Fermi pockets in the SDC of rhombohedral graphene, shown in the microscopic graphene BZ. Right inset shows the winding of the pairing phase arg[$\Delta(\mathbf{k})$] obtained by solving Eliashberg equations at displacement field $D = 53$ meV and carrier density $n_e = 0.56 \times 10^{12} cm^{-2}$.}
\label{fig:r5g}
\end{figure}

\paragraph{Rhombohedral graphene-}
We now consider a realistic Hamiltonian for rhombohedral pentalayer graphene~\cite{DongAHC2024,soejima2024anomalous,dong2024theory,Zhou2024} detailed in the SM~\cite{SM}, and focus on the SDC regime at low carrier density in the presence of a displacement field~\cite{Feng2026SDC,Dong2026MWC}. 
Here, an SDC shows up at the phase boundary between a topologically trivial Wigner crystal and a halo-anomalous Hall crystal in self-consistent Hartree-Fock numerics~\cite{Feng2026SDC}. 
Following an identical procedure to the $\lambda-$jellium model, we derive the electron-phonon coupling and the resultant electron-electron interaction in RMG. 
Then, solving the corresponding Eliashberg equation on an inversion-symmetrized Fermi surface (within a polarized valley), we find a chiral p-wave superconductor with Cooper pair angular momentum 1 and $C_{\rm BdG} = 1$ on the topologically trivial ($C = 0$) Wigner crystal side.
We note that our linearized Eliashberg equation does not yield a superconducting solution if the Fermi surface strongly breaks inversion. 

Although the physics is nearly identical to the previously studied $\lambda$-jellium model, there are a couple of key differences, arising from the fact that graphene has two valleys $K$ and $K^\prime$.
Here we assumed the Hamiltonian of the $K$ valley: starting with the time-reversal related $K^\prime$ valley instead simply changes the sign of $C_{\rm BdG}$. 
Further, since the Fermi pockets in graphene are either at the $K$ point or the $K^\prime$ point, the Cooper pair carries finite center-of-mass momentum [Fig.~\ref{fig:r5g}] and breaks translation symmetry at the microscopic graphene lattice scale, while preserving approximate translation symmetry on the crystal scale. 
Accordingly, the chiral superconducting state is actually a topological pair density wave.

\paragraph{Experimental signatures.-}
The superconductivity born out of an SDC carries its unique experimental fingerprints. 
Specifically, the $I$-$V$ curve should exhibit two nonlinear features at two different voltage scales, one associated with the de-pairing of the superconductor, and the other with the de-pinning of the underlying crystal. 
Within a simple two-fluid picture of this phenomenon, the superconducting condensate holds the voltage at zero up to a critical current, whereas the crystal holds its own current at zero up to a critical voltage, suggesting a peak in the differential resistance $dV/dI$ at the de-pairing scale and a dip at the de-pinning scale. 
Extrinsic factors like disorder can affect these scales and determine their relative order. 
We leave a full theoretical modeling of this phenomenon to future work, but note that some such features may have already been seen in certain samples of rhombohedral pentalayer graphene~\cite{Han2025Chiral}.


\paragraph{Conclusions and outlook-} 
We showed that phonons of electronic origin in self-doped electronic crystals can mediate pairing and lead to a chiral topological superconductor.
Our work demonstrates how strong electronic correlations, band topology and criticality stitch together to provide a new all-electronic route towards superconductivity from repulsive interactions, with potential applications to rhombohedral graphene. 
The phonon-mediated attraction may also provide a natural explanation to the two-dome structure of SC1 and SC2 in this system~\cite{han2025exactmodelschiralflatband}. 

More generally, our self-consistent computations indicate a novel physical regime of electron-phonon coupling (see also Ref.~\cite{froese2026influencequantumgeometryphase,kim2026stabilitycollectivemodesquasiparticlephonon}). 
The characteristic phonon frequency, although smaller, is of the same order of the Fermi energy (SM Fig.\ref{fig:ptoe}), since there is no parametric separation of masses between the itinerant charge carriers and the constituents of the crystal -- they are both electrons.
Therefore, our work showcases a platform to explore electron-phonon problems away from the usual adiabatic limit, where more exotic phases such as quantum spin liquids can be stabilized~\cite{PhysRevLett.130.186404,doi:10.1073/pnas.2426111122}. 
Further, our work also motivates careful theoretical and experimental studies of non-linear conductance in superconductors proximate to electronic crystals to probe the relevance of the crystalline phase to superconductivity. 

\begin{acknowledgments}
\paragraph{Acknowledgments--}
We thank F. Desrochers, T. Senthil, M. P. Zaletel and A. Vishwanath for inspiring conversations. 
S.Z. and S.C. are grateful to R. Sahay for a collaboration on a related project. 
We thank the authors of Ref.~\cite{toappear} for coordinating submission. Z.~H. is supported by the Gordon and Betty Moore Foundation EPiQS Award 8683.  Z.~D is funded by a grant from the Simons Foundation (651440, Ashvin Vishwanath) through the Simons Collaboration on Ultra-Quantum Matter.
\end{acknowledgments}

\paragraph{Note added--} Towards the completion of this work, we became aware of  papers~\cite{miao2026cooper,wang2026chiral} which consider different approaches to the problem of superconductivity proximate to a Wigner crystal.

\bibliography{refs}

\clearpage

\newpage

\onecolumngrid

\vspace{0.3cm}

\supplementarysection

\input{supp.tex}

\vfill

\end{document}

%% file: supp.tex
\makeatletter
\@addtoreset{subsection}{section}
\@addtoreset{subsubsection}{subsection}
\renewcommand{\thesubsection}{\thesection.\arabic{subsection}}
\renewcommand{\thesubsubsection}{\thesubsection.\arabic{subsubsection}}
\newcommand{\supp@heading}[3]{
  \par\addvspace{2.5ex \@plus 1ex \@minus .2ex}%
  \refstepcounter{#1}%
  {\centering #2\csname the#1\endcsname\quad #3\par}\nobreak
  \addvspace{1ex \@plus .2ex}\@afterheading}
\renewcommand{\section}[1]{\supp@heading{section}{\normalfont\large\bfseries}{#1}}
\renewcommand{\subsection}[1]{\supp@heading{subsection}{\normalfont\normalsize\bfseries}{#1}}
\renewcommand{\subsubsection}[1]{\supp@heading{subsubsection}{\normalfont\normalsize\itshape}{#1}}
\makeatother

\begin{center}
  {\normalfont\large\bfseries Supplemental Material for\\[0.5ex]
   ``Goldstone-Mediated Chiral Superconductivity in Self-Doped Electron Crystals''\par}
  \vspace{2.5ex}
  {\normalfont\normalsize Shuchen Zhang,$^{1}$ Zhihuan Dong,$^{2}$ Zhaoyu Han,$^{2}$ and
   Shubhayu Chatterjee$^{1}$\par}
  \vspace{1.5ex}
  {\normalfont\small\itshape
   $^{1}$Department of Physics, Carnegie Mellon University, Pittsburgh, Pennsylvania 15213, USA\\
   $^{2}$Department of Physics, Harvard University, Cambridge, Massachusetts 02138, USA\par}
\end{center}
\vspace{2ex}

\section{Units and parameters}
We list the units and parameters used in our numerics here.

\begin{itemize}
\item Length: $a_\rho = r_s \epsilon \frac{m}{m^*} a_B$, where $\epsilon$ is the dielectric constant and $a_B$ is the Bohr radius.  The electron density is $n_e = 1/(\pi a_\rho^2)$. Momenta are in $a_\rho^{-1}$.
\item Energy: $E_{LJ}\text{(code unit)} = \hbar^2/(m^* a_\rho^2) = \mathrm{Ha}^*/r_s^2 = 2\,\mathrm{Ry}^*/r_s^2$.
  The kinetic energy is $p^2/2m^*$ and the Coulomb scale is $e^2/(\epsilon a_\rho) = r_s\,E_{LJ}$.
\item Temperature: $k_B T$ in units of $E_{LJ}$. For $m^*=m_e$, $\epsilon=1$:
  $E_{LJ}/k_B = 3.158\times10^5\,\mathrm K/r_s^2$, i.e.\ $790$\,K at $r_s=20$
  (scale by $(m^*/m_e)/\epsilon^2$ otherwise). 
\item Gate distance $d_g$ is in units of $a_\rho$ 
\end{itemize}

\section{Self-consistent Hartree-Fock}
For most of our work, we use the same model and parameters for the $\lambda$-jellium model as in Ref.~\cite{Feng2026SDC}. Therefore, we only list the differences here: 

1. We set the gate distance to be at a fixed inter-particle distance 5, which makes the changes in our paper a function of only $\lambda$ and $r_s$. Feng et al. fix the physical gate distance, which makes the screening length change as the inter-particle distance changes. 

2. We cut off the extended Brillouin zone at $\pm$ 2.5 reciprocal vectors, but Feng et al. cut off at 6. 

3. We consider screening from the mobile carriers in the metallic state. After the Hartree-Fock with the bare interactions as in Feng et. al, we compute the static polarization function of the lowest HF band with RPA approximation
\begin{equation}
    \Pi(\mathbf{q}) = \frac{1}{N^2} \sum \limits_{\mathbf{k}} \braket{\psi_\mathbf{k} |e^{-i\mathbf{q} \cdot \mathbf{r}}| \psi_{\mathbf{k} + \mathbf{q}}}^2 \frac{f_{\mathbf{k}} - f_{\mathbf{k} + \mathbf{q}}}{E_{\mathbf{k} + \mathbf{q}} - E_\mathbf{k}},
\end{equation} at T = $\frac{1}{2} \frac{\delta k}{N} v_{Fmed}$, where f is the Fermi-Dirac function, $\delta k$ is the momentum grid spacing, $v_{Fmed}$ is the median of the Fermi velocity of the HF band 0 state computed at the nearest neighbors of a point on the Fermi surface, and the form factor is summed over all the reciprocal vectors. At small $\mathbf{q}$, the difference becomes a derivative. We compute $\Pi$ to $\mathbf{q}_c = 2.5 p_F$, and then approximate the tail with $\Pi(\mathbf{q}_c) (\frac{q_c}{q})^2$. 

Then, we replace $V(\mathbf{q})$ in the exchange interaction by the screened \begin{equation}
    W(\mathbf{q}) = \frac{V(\mathbf{q})}{1 + \Pi(\mathbf{q})V(\mathbf{q})} \, .
\end{equation} Note that all the other interactions are still bare to avoid over-counting. They will be screened later in the phonon computation.

With these, we obtain the self-doped crystal state for further studies.

\section{Obtaining the collective mode spectra and identifying phonon modes}
We follow the formalism developed in Ref.~\cite{Vasiliou2024Goldstone} to compute the collective modes by solving a Bethe Saltpeter equation, which is a generalized eigenvalue equation that is equivalent to time-dependent Hartree-Fock as discussed in Ref.~\cite{Vasiliou2024Goldstone} .
We define the Bloch states to be 
\begin{eqnarray} 
|\psi_{\alpha}(\mathbf{k})\rangle=&&e^{i\mathbf{k}\cdot\mathbf{r}}|u_{\alpha}(\mathbf{k})\rangle,\\
|u_{\alpha(\mathbf{k})}\rangle=&&\sum_{\mathbf{G}'} c_{\alpha}(\mathbf{k},\mathbf{G}')\,e^{i\mathbf{G}'\cdot\mathbf{r}}\,
|\chi_\lambda(\mathbf{k}+\mathbf{G}')\rangle,\\
|\chi_\lambda(\mathbf{P})\rangle=&&\frac{1}{\sqrt{1+\lambda^2|\mathbf{P}|^2}}
\begin{pmatrix}1\\ \lambda(P_x+iP_y)\end{pmatrix},
\end{eqnarray} where $\alpha$ labels the HF bands and $\chi$ is the eigenstate of the $\lambda-$jellium model. Then, we define the form factors 
\begin{eqnarray}
    \Lambda_{\alpha\beta}(\mathbf{Q},\mathbf{k})\equiv&& \langle u_{\alpha}(\mathbf{k})|u_{\beta}(\mathbf{k}+\mathbf{Q})\rangle 
\end{eqnarray}.

The generalized eigenvalue equation to determine the collective mode spectrum is given by~\cite{Vasiliou2024Goldstone}
\begin{eqnarray} &&\sum_{\mathbf{k}'\sigma'}A(\mathbf{q})_{\mathbf{k}\sigma,\mathbf{k}'\sigma'}\,\phi_{\sigma'}(\mathbf{q},\mathbf{k}')
=\omega_{\mathbf{q}s}\,\eta_{\mathbf{q}s}\sum_{\mathbf{k}'\sigma'}B(\mathbf{q})_{\mathbf{k}\sigma,\mathbf{k}'\sigma'}\,
\phi_{\sigma'}(\mathbf{q},\mathbf{k}'),
 \end{eqnarray}
\begin{equation}
B(\mathbf{q})_{\mathbf{k}\sigma,\mathbf{k}'\sigma'}=\big(n_\alpha(\mathbf{k})-n_\beta(\mathbf{k}-\mathbf{q})\big)\,
\delta_{\alpha\mu}\delta_{\beta\nu}\delta_{\mathbf{k}\mathbf{k}'},
\end{equation}
\begin{equation}
A(\mathbf{q})_{\mathbf{k}\sigma,\mathbf{k}'\sigma'}=\big|E_\alpha(\mathbf{k})-E_\beta(\mathbf{k}-\mathbf{q})\big|\,
\delta_{\alpha\mu}\delta_{\beta\nu}\delta_{\mathbf{k}\mathbf{k}'}
+K(\mathbf{q})_{\mathbf{k}\sigma,\mathbf{k}'\sigma'},
\end{equation}
\begin{equation}
    K(\mathbf{q})=K^H(\mathbf{q})-K^F(\mathbf{q}),
\end{equation}
where $\sigma = (\alpha, \beta)$, $\phi$ is the wave function of a collective mode, $n$ is the occupation number of an HF state, E is the HF energy with W in the exchange interaction, 
\begin{eqnarray}
&&K^H(\mathbf{q})_{\mathbf{k}\alpha\beta,\mathbf{k}'\mu\nu}
=\sum_{\mathbf{G}}V(\mathbf{q}+\mathbf{G})\,
\Lambda^*_{\beta\alpha}(\mathbf{q}+\mathbf{G},\mathbf{k}-\mathbf{q})\,
\Lambda_{\nu\mu}(\mathbf{q}+\mathbf{G},\mathbf{k}'-\mathbf{q}),
\end{eqnarray}
\begin{eqnarray}  &&K^{F}(\mathbf{q})_{\mathbf{k}\alpha\beta,\mathbf{k}'\mu\nu}
= \sum_{\mathbf{G}}W(\mathbf{k}-\mathbf{k}'+\mathbf{G})\,
\Lambda_{\alpha\mu}(\mathbf{k}'-\mathbf{k}+\mathbf{G},\mathbf{k})\,
\Lambda^*_{\beta\nu}(\mathbf{k}'-\mathbf{k}+\mathbf{G},\mathbf{k}-\mathbf{q})
\end{eqnarray} Note that the exchange interaction here is also the screened W. 

Now we explain how we select the phonons from the solution since they are mixed with the particle-hole continuum. The naive method of selecting the two modes with the lowest energies does not work  well. 

Phonons are really the strongest responses to a displacement, so we apply a small displacement and see the responses of all the modes. A mode's response is proportional to its overlap with the displacement and inversely proportional to its frequency. This idea leads to \ref{eq:D} which we use to select the phonons. We expect this to be correct in most cases. At large q, it is possible that the response is shared by several modes and there is not a single well-defined phonon mode. In our program, this criterion works well in the range of momentum we care about. Now we derive Eq.\ref{eq:D}.

\subsection{Momentum density operator}
The unitary operator corresponding to a displacement field to the first order is $\mathbf{u}(\mathbf{r})$ is $U[\mathbf{u}]=\exp\!\big(-i\int d^2r\,\mathbf{u}(\mathbf{r})\cdot\hat{\bo g}(\mathbf{r})\big)$, where the
momentum density \begin{equation}
\hat{\bo g}(\rr)=\tfrac12\big[\psi^\dagger(-i\nabla\psi)+(i\nabla\psi^\dagger)\psi\big].
\end{equation}. In the mode basis and the momentum space, it is \begin{equation}
\hat Q_\nu(\mathbf{q})\equiv\int d^2r\,e^{-i\q\cdot\rr}\,\eh_\nu\cdot\hat{\bo g}(\rr).
\end{equation}
For a wave function $\psi(\rr)=\mathcal{A}^{-1/2}\sum_\PP e^{i\PP\cdot\rr}c_\PP$,
\begin{align}
\int e^{-i\q\cdot\rr}\psi^\dagger(-i\nabla\psi)&=\sum_{\PP\PP'}\PP'\,c^\dagger_\PP c_{\PP'}\,\delta_{\PP',\PP+\q}\nonumber \\
&=\sum_\PP(\PP+\q)\,c^\dagger_\PP c_{\PP+\q}, \\
\int e^{-i\q\cdot\rr}(i\nabla\psi^\dagger)\psi&=\sum_\PP\PP\,c^\dagger_\PP c_{\PP+\q},
\end{align}
so
\begin{equation}
\hat Q_\nu(\q)=\sum_{\PP}\Big(\PP+\tfrac{\q}{2}\Big)\!\cdot\eh_\nu\;c^\dagger_{\PP}c_{\PP+\q}
, \hat Q_\nu(0)=\hat P_\nu\ .
\label{eq:Qop}
\end{equation}
In the lowest band of the $\lambda-$jellium model, this operator is \begin{equation}
\hat Q_\nu(\q)=\sum_\PP S(\PP,\PP+\q)\Big(\PP+\tfrac\q2\Big)\!\cdot\eh_\nu\;a^\dagger_\PP a_{\PP+\q},
\label{eq:Qproj}
\end{equation} $$S(\PP,\PP')=\braket{\chi_\lambda(\PP)|\chi_\lambda(\PP')},$$ where $\chi$ is the eigenstate of the band. At $\q=0$, $S=1$ and $\hat Q_\nu(0)=\sum_\PP P_\nu\,a^\dagger_\PP a_\PP$ is the exact generator of translations of
the projected model. At non-zero $\q$ the projected form is a choice. Any other choice differs by $O(q)$. In our program, we compute the matrix elements of Eq.\ref{eq:Qproj} in the HF basis and the longitudinal and transverse mode basis\begin{equation}
\eh_L=\hat\q,\qquad \eh_T=\hat z\times\hat\q,
\label{eq:pol}
\end{equation}

\subsection{Time-dependent Hartree-Fock with an external field}
Time-dependent Hartree--Fock (TDHF) evolves the density matrix with the self-consistent HF Hamiltonian plus any
external one-body field $h_{\rm ext}(\rt)$:
\begin{equation}
i\,\partial_\rt\Rm=\big[h[\Rm]+h_{\rm ext}(\rt),\,\Rm\big].
\end{equation}
Write $\Rm=\Rm^0+\delta\Rm$ with $\delta\Rm$ small, so that $h[\Rm]=h[\Rm^0]+\delta h$ with $\delta h$ linear in
$\delta\Rm$. Using $[h[\Rm^0],\Rm^0]=0$ and keeping first-order terms,
\begin{equation}
i\,\partial_\rt\delta\Rm=\big[h[\Rm^0],\delta\Rm\big]+\big[\delta h+h_{\rm ext},\Rm^0\big].
\end{equation}
Both $h[\Rm^0]$ and $\Rm^0$ are diagonal, so for the matrix element between $\ell$ and $\ell'$
\begin{equation}
i\,\partial_\rt\delta\Rm_{\ell\ell'}=(E_\ell-E_{\ell'})\,\delta\Rm_{\ell\ell'}+(n_{\ell'}-n_\ell)\,\big(\delta h_{\ell\ell'}+(h_{\rm ext})_{\ell\ell'}\big).
\label{eq:tdhf}
\end{equation}

If $n_\ell=n_{\ell'}$ the second term is absent and nothing drives $\delta\Rm_{\ell\ell'}$ at linear order: only elements between an
occupied and an empty state move. 

Take $\ell=(\kk-\q,\beta)$, $\ell'=(\kk,\alpha)$ and define
\begin{equation}
\xi_{\kk\sigma}(\rt)\equiv\delta\Rm_{(\kk-\q\,\beta),(\kk\alpha)}
=\delta\big\langle\hat\gamma^\dagger_\alpha(\kk)\,\hat\gamma_\beta(\kk-\q)\big\rangle .
\label{eq:xi}
\end{equation}
It is the amplitude of the particle--hole pair $(\kk,\sigma)$ in the perturbed state. For this choice
$n_{\ell'}-n_\ell=B_{\kk\sigma}$ and $E_\ell-E_{\ell'}=B_{\kk\sigma}|E_\alpha(\kk)-E_\beta(\kk-\q)|$.
The change of the HF potential is linear in the pair amplitudes; its kernel is, by definition, the $K(\q)$ of the
draft:
$\delta h_{(\kk-\q\,\beta),(\kk\alpha)}=\sum_{\kk'\sigma'}K(\q)_{\kk\sigma,\kk'\sigma'}\,\xi_{\kk'\sigma'}$.
Crystal momentum is conserved, so pairs at transfer $\q$ couple only to each other. Collecting the terms,
Eq.~\eqref{eq:tdhf} becomes
\begin{align}
\;i\,\partial_\rt\xi=B(\q)\big[A(\q)\,\xi+\mathcal{O}^{\rm ext}\big],  \\
\mathcal{O}^{\rm ext}_{\kk\sigma}(\rt)=\big\langle\psi_\beta(\kk-\q)\big|h_{\rm ext}(\rt)\big|\psi_\alpha(\kk)\big\rangle\; \nonumber
\label{eq:eom} 
\end{align}

\subsection{Linear response theory}
Let $\Op$ be any one-body operator that lowers the crystal momentum by $\q$ (as
$\hat Q_\nu(\q)$ does), and drive the system with
\begin{equation}
h_{\rm ext}(\rt)=\zeta\big(\Op\,e^{-i\Omega\rt}+\Op^\dagger e^{i\Omega\rt}\big),
\end{equation}
where $\zeta$ is a small real amplitude and $\Omega$ the driving frequency. The pair vector of the probe is
\begin{equation}
\mathcal{O}_{\kk\sigma}\equiv\big\langle\psi_\beta(\kk-\q)\big|\Op\big|\psi_\alpha(\kk)\big\rangle .
\label{eq:Ovec}
\end{equation}

In the steady state $\xi(\rt)=\bar\xi\,e^{-i\Omega\rt}$, and
Eq.~\eqref{eq:eom} becomes $\Omega\bar\xi=B(\q)A(\q)\bar\xi+\zeta B(\q)\mathcal{O}$, i.e.
\begin{equation}
\bar\xi=\zeta\,(\Omega-B(\q)A(\q))^{-1}B(\q)\,\mathcal{O}.
\end{equation}
Expand the inverse in the modes
\begin{equation}
(\Omega-B(\q)A(\q))^{-1}=\sum_s\frac{\eta_{\q s}\,\phi^s\,\phi^{s\dagger}B(\q)}{\Omega-\eta_{\q s}\omega_{\q s}}
\label{eq:resolvent}
\end{equation}
With $B(\q)^2=1$,
\begin{equation}
\bar\xi=\zeta\sum_s\frac{\eta_{\q s}\,\phi^s}{\Omega-\eta_{\q s}\omega_{\q s}}
\Big(\sum_{\kk\sigma}\big[\phi^s_\sigma(\q,\kk)\big]^*\mathcal{O}_{\kk\sigma}\Big).
\end{equation}
The drive excites mode $s$ in proportion to the (Euclidean) overlap of the mode with the probe vector, and
resonantly when $\Omega$ approaches the mode frequency $\eta_{\q s}\omega_{\q s}$.

The response function is
\begin{equation}
\Xc_{\mathcal O}(\q,\Omega)\equiv\frac{\delta\langle\Op^\dagger\rangle}{\zeta\,e^{-i\Omega\rt}}
=\sum_{\kk\sigma}\mathcal{O}^*_{\kk\sigma}\big[(\Omega-B(\q)A(\q))^{-1}B(\q)\mathcal{O}\big]_{\kk\sigma},
\end{equation}
and inserting Eq.~\eqref{eq:resolvent},
\begin{equation}
\;\Xc_{\mathcal O}(\q,\Omega)=\sum_s\frac{\eta_{\q s}\,
\big|\sum_{\kk\sigma}[\phi^s_\sigma(\q,\kk)]^*\,\mathcal{O}_{\kk\sigma}\big|^2}{\Omega-\eta_{\q s}\,\omega_{\q s}}\;
\label{eq:chi}
\end{equation}
This is the Lehmann representation of the mode's response.

A time-independent push corresponds to $\Omega=0$:
\begin{equation}
\Xc_{\mathcal O}(\q,0)=-\sum_s\frac{\big|\sum_{\kk\sigma}[\phi^s_\sigma(\q,\kk)]^*\,\mathcal{O}_{\kk\sigma}\big|^2}
{\omega_{\q s}}
\label{eq:static}
\end{equation}
When we displace the electrons by $\uu(\rr)=\zeta\,\eh_\nu\big(e^{-i\q\cdot\rr}+e^{i\q\cdot\rr}\big)$ with the unitary operator
$U[\uu]=\exp\!\big[-i\zeta\big(\hat Q_\nu(\q)+\hat Q_\nu(\q)^\dagger\big)\big]$, and the HF density matrix changes
to first order by $\delta\Rm=-i\zeta\,\big[Q_\nu(\q)+Q_\nu(\q)^\dagger,\Rm^0\big]$ ($Q_\nu(\q)$ is the
matrix of $\hat Q_\nu(\q)$ in the HF basis). Its pair components at transfer $\q$ are
\begin{equation}
\xi^{\rm disp}_{\kk\sigma}=-i\zeta\,\big(n_\alpha(\kk)-n_\beta(\kk-\q)\big)
\big\langle\psi_\beta(\kk-\q)\big|\hat Q_\nu(\q)\big|\psi_\alpha(\kk)\big\rangle
\end{equation}
We can define the pair vector of the displacement
\begin{equation}
\;\Yc_{\nu,\kk\sigma}(\q)=B_{\kk\sigma}\,
\big\langle\psi_\beta(\kk-\q)\big|\hat Q_\nu(\q)\big|\psi_\alpha(\kk)\big\rangle\;
\label{eq:Y}
\end{equation}
Its static response function, Eq.~\eqref{eq:static} with $\mathcal{O}=\Yc_\nu(\q)$, is therefore how far the
crystal is displaced per unit static push along the displacement:
\begin{equation}
-\Xc_{\Yc_\nu}(\q,0)=\sum_s\frac{\big|\sum_{\kk\sigma}[\phi^s_\sigma(\q,\kk)]^*\,\Yc_{\nu,\kk\sigma}(\q)\big|^2}{\omega_{\q s}} .
\label{eq:chiY}
\end{equation}
We normalize this and get  \begin{equation}
D_{s,\nu}(\mathbf{q})=
\frac{\big|\sum_t \phi_{s,t}\,B_t\,Q_{\nu,t}(\mathbf{q})\big|^2}{2\omega_s}.
\label{eq:D}
\end{equation}
The two modes with the highest D are the phonons we need. Sometimes two modes can have similar responses, and one needs to manually check the dispersion to tell which is the correct phonon mode. We show an example of the modes in Fig.\ref{fig:results} (a).

\section{Electron-phonon coupling and pairing interaction}
We compute the electron-Goldstone coupling g,f for these two modes defined in the following action
\begin{eqnarray}
S_{el-ph}=\int_{0}^{1/T} d\tau\frac{1}{\sqrt{N}}\sum_{\mathbf{q}} \sum_{\mathbf{k}} \overline{\psi}_{\mathbf{k},\alpha}\psi_{\mathbf{k}-\mathbf{q},\beta} 
\sum_{s= L, T}(g_{\mathbf{q},\alpha\beta}^{s}(\mathbf{k})\phi_{-\mathbf{q},s} + f_{\mathbf{q},\alpha\beta}^{s}(\mathbf{k})i\partial_{\tau}\phi_{-\mathbf{q},s}).
\end{eqnarray} 
We use the same method as Ref.~\cite{Vasiliou2024Goldstone} to compute the vertices $g^s_{\mathbf{q}}$ and $f^s_{\mathbf{q}}$. The only difference is that we use the screened HF states and screen the exchange interactions.
Next, we discuss how we can use the vertices to find the pair interaction by integrating out the Goldstone modes.

Our pairing interaction in the BCS channel (at zero COM momentum) can be written as
\begin{equation}
  \;
  V(i\nu; \mathbf k, \mathbf k') = V_C(i\nu;\kk,\kk') + V_F(\mathbf k,\mathbf k')
  + V_G(i\nu; \mathbf k, \mathbf k') \;
  \label{eq:Vnu}.
\end{equation}
We explain how we construct it here.

All the terms are first computed between mesh points and then interpolated onto the Fermi surface. The Coulomb term contains the bare repulsion and its retarded screening by all the
collective modes other than the two phonons,
\begin{equation}
  V_C(i\nu;\kk,\kk')=V_C^{0}(\kk,\kk')+V_{\rm rest}(i\nu;\kk,\kk'),
  \label{eq:VCscr}
\end{equation}. 
$V_C^0$, the Hartree legs of the vertices and the legs of $V_{\rm rest}$ use the bare $V$, because the collective modes themselves generate the screening of these lines. The screened V only appears in exchange interactions.

Projected onto the lowest HF band, the density operator is
\begin{equation}
  \hat\rho(\mathbf Q)=\sum_{\kk}\Lambda_{00}(\mathbf Q,\kk)\,\hat\gamma^\dagger_0(\kk)\,\hat\gamma_0(\kk+\mathbf Q).
\end{equation}
Collecting the terms $\hat\gamma^\dagger_0(\kk)\hat\gamma^\dagger_0(-\kk)\hat\gamma_0(-\kk')\hat\gamma_0(\kk')$ of
$\frac12\sum_{\mathbf Q}V(\mathbf Q)\hat\rho(\mathbf Q)\hat\rho(-\mathbf Q)$ gives
\begin{equation}
\begin{split}
  V_C^0(\kk,\kk')=\sum_{\G}V(\kk'-\kk+\G)\,&\Lambda_{00}(\kk'-\kk+\G,\kk)\\
  &\Lambda_{00}(\kk-\kk'-\G,-\kk)
\end{split}
  \label{eq:VCapp}
\end{equation}
Note that the two form factors are not equal because time-reversal symmetry is broken.

The remaining collective modes screen the Coulomb repulsion. Within TDHF the dynamically screened Coulomb interaction is
\begin{equation}
  V_{\rm scr}(\q,i\nu)=V+V\chi(\q,i\nu)V ,
  \label{eq:Wscr}
\end{equation}
where $\chi$ is the density response of the HF state. The bare $V$ couples an electron to the density
it creates, and $\chi$ is the response of the medium to that density. The medium is described by the same
generalized eigenvalue equation as the phonons. By Eq.~\eqref{eq:resolvent}, any response function at
transfer $\q$ is therefore a sum of simple poles, one for each solution $\phi^n(\q)$, located at its
eigenvalue $\eta_n\omega_n$.

For the Cooper channel, the probe is the potential created by the transition of one electron. The leg $\kk'\to\kk$ drives the medium and the leg $-\kk'\to-\kk$ feels its
response. With two different pair vectors, the same steps that lead to Eq.~\eqref{eq:chi} give a sum of
poles whose numerator is the product of the two overlaps. The residues are
\begin{equation}
  R^n(\kk,\kk')=\big[h_{\q}(\kk)\cdot\phi^n(\q)\big]\,\big[h_{\q}(-\kk')\cdot\phi^n(\q)\big]^*,
\end{equation}
where $h\cdot\phi=\sum_{\kk''\sigma''}h_{\kk''\sigma''}\,\phi_{\sigma''}(\q,\kk'')$. The second leg, $h_{\q}(-\kk')$,
is the transition $-\kk\to-\kk'$; its complex conjugate is the Cooper leg $-\kk'\to-\kk$.

$R^n$ is the product of the couplings of mode $n$ to the two electrons. Each screening mode is therefore
exchanged in exactly the same way as a phonon: it is emitted on one leg and absorbed on the other. The
residue says how strongly mode $n$ couples to the pair, and the pole says at which frequency it responds. The
interaction induced through the full TDHF response is
\begin{equation}
  V_{\rm ind}(i\nu;\kk,\kk')=\sum_n\frac{\eta_n\,R^n(\kk,\kk')}{i\nu-\eta_n\omega_n}.
  \label{eq:Vind}
\end{equation}
At $\nu=0$ it equals $-h_{\q}(\kk)\,A(\q)^{-1}h_{\q}(-\kk')^\dagger$. The sum $V_C^0+V_{\rm ind}$ is the
Cooper-channel matrix element of Eq.~\eqref{eq:Wscr}.

The two phonons are already in $V_F+V_G$ with their full vertices, so we remove their poles:
\begin{equation}
  V_{\rm rest}(i\nu;\kk,\kk')=\sum_{n\notin\{s\pm\}}\frac{\eta_n\,R^n(\kk,\kk')}{i\nu-\eta_n\omega_n}.
  \label{eq:Vrest}
\end{equation}
The excluded poles are $+\omega_{\q s}$ and $-\omega_{-\q s}$ of the two Goldstone branches. This is the
screening part of $V_C$ in Eq.~\ref{eq:VCscr}.

The legs of $V_{\rm rest}$ are density (Hartree) legs only. The exchange part of an external leg is a vertex
correction, the external electron exchanging with the medium, rather than screening.

We define the following functions
\begin{equation}
\begin{split}
  \mathcal A_s&=g^{s}_{\q}(\kk)\,g^{s}_{-\q}(-\kk),\\
  \mathcal B_s&=g^{s}_{\q}(\kk)\,f^{s}_{-\q}(-\kk)-f^{s}_{\q}(\kk)\,g^{s}_{-\q}(-\kk),\\
  \mathcal C_s&=f^{s}_{\q}(\kk)\,f^{s}_{-\q}(-\kk).
\end{split}
\end{equation}
Then, with $c_s=\omega_s/|\q|$, integrating out the two Goldstone fields of $S_{el-B}$ for the two gapless phonon modes with the boson propagator gives
\begin{align}
  V_F(\kk,\kk')&=-\frac1N\sum_s\frac{c_s}{a w_s}\,\mathcal C_s,\\
  V_G(i\nu;\kk,\kk')&=-\frac1N\sum_s\frac{c_s}{a w_s}\,
  \frac{\mathcal A_s+\nu\,\mathcal B_s-\nu^2\,\mathcal C_s}{\nu^2+\omega_s^2},
\end{align} where $w_s$ is a constant introduced in \cite{Vasiliou2024Goldstone}.
These are the $V_F$ and $V_G$ of Eq.~\ref{eq:Vnu}.

We characterize the electron-phonon coupling strength by the McMillan's dimensionless parameter, slightly generalized here for finite winding $w$ on Fermi surface and plotted in the main text Fig.\ref{fig:results}(c). It is defined as: 
\begin{eqnarray} \lambda_w = 2\int_0^\infty\!d\Omega\,\frac{\alpha^2F_w(\Omega)}{\Omega},  
\end{eqnarray}
\begin{eqnarray}
     & \alpha^2F_w(\Omega)=\frac{1}{N(0)}\oint_{S(\mathbf{k})} \oint_{S(\mathbf{k}^\prime)} \sum_s
  \frac{\kappa_s(\mathbf q)}{N(0)}{\rm Re}\Big\{\Delta_w^*(\mathbf k)\Delta_w(\mathbf k')
  \big[g_s(\mathbf k,\mathbf q)g_s(-\mathbf k,-\mathbf q)+\omega_s^2 f_s(\mathbf k,\mathbf q)f_s(-\mathbf k,-\mathbf q)\big]\Big\}\,\delta\big(\Omega-\omega_s(\mathbf q)\big), \nonumber \\
\end{eqnarray}
where we have defined 
\begin{equation}
   \oint_{S(\mathbf{k})} =  \oint \frac{dS_{\mathbf k}}{(2\pi)^2\,v_F(\mathbf k)}, \; 
  \Delta_w(\mathbf k)=e^{iw\theta_{\mathbf k}}e^{-i(\beta_{\mathbf k}+\beta_{-\mathbf k})}, \kappa_s(\mathbf q)=\frac{1}{2|\mathbf q|\,a\,w_s}
\end{equation} where s is a Goldstone mode, $\theta$ is the Fermi surface polar angle, and $\beta_{\mathbf{k}}$ is the gauge we choose. $\alpha^2 F$ can be thought of the spectral density of the pairing interaction in an angular momentum channel weighted by the density of states.



\section{Eliashberg equations}
\label{sec:eliash}
We solve the Eliashberg equation 
\begin{equation}
    \lambda \Delta(i\omega, \mathbf{k}) = -\frac{1}{N}T \sum_{i\omega'}\sum_{\mathbf{k}'} \frac{V(i\omega - i\omega', \mathbf{k}'-\mathbf{k})\Delta(i\omega', \mathbf{k}')}{\omega^{'2}Z^2(i\omega', \mathbf{k}') + (E_k - \mu)^2} 
    \label{eq:Eliash}
\end{equation} to study superconductivity, where $\Delta$ is Z times the gap function, \begin{equation}
    Z(i\omega_n) = 1 + \frac{\pi T}{\omega_n} \sum_m \lambda_Z(\omega_n - \omega_m) sgn(\omega_m),
\end{equation}
\begin{equation}
        \lambda_Z(i\nu) =  \oint_{FS} \frac{1}{N_0 (2\pi)^2 |v_\mathbf{k}|} dl_\mathbf{k}   \oint_{FS} -P(\mathbf{k}, \mathbf{k}', i\nu) \frac{1}{(2\pi)^2 |v_{\mathbf{k}'}|} dl_{\mathbf{k}'} 
        \label{eq:lambdaz}
\end{equation}, $N_0$ is the density of states at the Fermi level, $v_\mathbf{k}$ is the group velocity in the lowest HF band, P contains the poles from the phonons and the retarded screened Coulomb repulsion, and with Eq.\ref{eq:Vnu} for V, by linearizing it on the Fermi surface from self-doping and projecting it to the lowest HF band. We include frequency dependency because the Fermi velocity is roughly 10 times larger than the phonon velocity, so retardation needs to be considered. This can also be seen from the following figure showing $\frac{\omega_L (2k_F)}{E_F}$.

\begin{figure}[!t]
\includegraphics[width=0.5\textwidth]{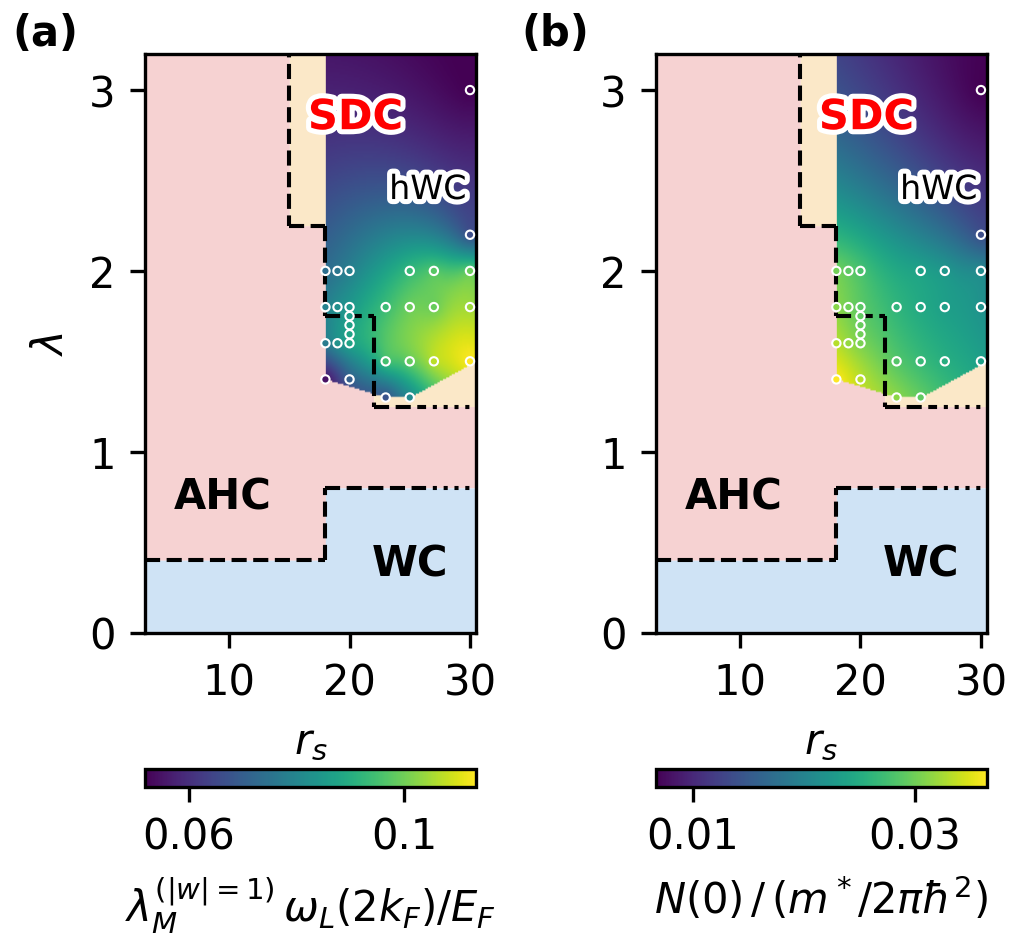}
\caption{(a)The Migdal parameter $\lambda_M \frac{\omega(2k_F)}{E_F}$ and (b)density of states at the Fermi level as a function of $\lambda$ and $r_s$. The interpolation rule is the same as main text Fig.2.}
\label{fig:ptoe}
\end{figure}

We provide a marginal justification to our treatment. The ratio of electron-phonon interaction energy to the Hartree-Fock band gap between the lowest band and the second-lowest band on average is 10 in the whole Brillouin zone. However, if we only look at the interaction energy with the momentum transfers on the Fermi surface, then the ratio drops to about 0.5. In addition, the Fermi energy is about the same as the energy difference from the Fermi level to the second-lowest band bottom. Therefore, this approximation is marginally OK. The $\Delta$ in the equation is the anomalous self-energy, i.e.\ $Z$ times the gap, and the eigenvector we report is this quantity.

The Fermi surface is the Luttinger contour of the lowest HF band, sampled at $n_{\rm FS}$ points $\kk_i$ at equally spaced angles ($n_{\rm FS}=24$ unless stated otherwise). Each point represents an arc of length $\delta l_i$. The momentum sum  splits into a sum over the contour and an energy integral,
\begin{equation}
  \frac1N\sum_{\kk'}F(\kk')\;\to\;\sum_j m_j^2\int_{\xi_{\rm min}}^{\xi_{\rm max}}d\xi\,F(\kk_j,\xi),
  m_j^2=\frac{A_c\,\delta l_j}{(2\pi)^2|v_{\kk_j}|},
  \label{eq:fsmeas}
\end{equation}
where $A_c$ is the area of the crystal unit cell and $v_{\kk}$ the group velocity of the lowest HF band. The
energy integral runs over the whole band, $\xi_{\rm min}=\min_{\kk}E_{\kk}-\mu$ and $\xi_{\rm max}=\max_{\kk}E_{\kk}-\mu$. It is done analytically:
\begin{equation}
  I(\omega)=\int_{\xi_{\rm min}}^{\xi_{\rm max}}\frac{d\xi}{\omega^2+\xi^2}
  =\frac{1}{|\omega|}\Big[\arctan\frac{\xi_{\rm max}}{|\omega|}+\arctan\frac{|\xi_{\rm min}|}{|\omega|}\Big].
  \label{eq:Ixi}
\end{equation}
In the denominator $\omega'^2Z^2+\xi^2$ we use $\int d\xi\,(\omega'^2Z^2+\xi^2)^{-1} \approx I(\omega')/Z$, i.e.\ we
neglect $Z$ inside arctan, which only matters when $|\omega'|Z$ is comparable to the band width, far above the frequencies that set $T_c$.

With $\phi_i(i\omega_n)=m_i\Delta(i\omega_n,\kk_i)$ the gap equation becomes the matrix eigenvalue problem
\begin{equation}
\begin{split}
  \lambda\,\phi_i(i\omega_n)=-T\sum_{m}\sum_{j}\,&m_i\,V(i\omega_n-i\omega_m;\kk_i,\kk_j)\,m_j\\
  &\times\frac{I(\omega_m)}{Z(i\omega_m)}\,\phi_j(i\omega_m),
\end{split}
  \label{eq:eliashdisc}
\end{equation}
with fermionic frequencies $\omega_n=(2n+1)\pi T$, $n=-N_\omega,\dots,N_\omega-1$. The cutoff is
$(2N_\omega+1)\pi T\geq\omega_{\rm max}=10\max(|\xi_{\rm min}|,\xi_{\rm max})$, so the number of frequencies grows
as $1/T$. $T_c$ is the temperature at which the largest eigenvalue is $\lambda=1$.

The electrons are spin polarized, so the gap must be odd, $\Delta(\kk)=-\Delta(-\kk)$. Let $\Pi$ be the permutation
that maps each contour point $\kk_i$ to the contour point at $-\kk_i$. We apply the projector
\begin{equation}
  P_{\rm odd}=\tfrac12\,(1-\Pi)
\end{equation}
to the pairing interaction, so only odd-parity solutions are found.

\subsection{Mass renormalization}
We explain how the Z factor is computed here. $Z$ is computed once per temperature. With the measure of Eq.~\eqref{eq:fsmeas},
\begin{equation}
  \lambda_Z(i\nu)=-\frac{\sum_{ij}m_i^2\,m_j^2\,P(i\nu;\kk_i,\kk_j)}{\sum_i m_i^2},
\end{equation}
$P=P_G+P_{\rm rest}$. $P_G$ is the normal-channel counterpart of the retarded part of $V$
\begin{equation}
\begin{split}
  P_G(i\nu;\kk,\kk')&=-\frac1N\sum_s\frac{c_s}{a w_s}\,
  \frac{\mathcal A^N_s+\nu\,\mathcal B^N_s-\nu^2\,\mathcal C^N_s}{\nu^2+\omega_s^2},\\
  \mathcal A^N_s&=g^{s}_{\q}(\kk)\,g^{s}_{-\q}(\kk'),\\
  \mathcal B^N_s&=g^{s}_{\q}(\kk)\,f^{s}_{-\q}(\kk')-f^{s}_{\q}(\kk)\,g^{s}_{-\q}(\kk'),\\
  \mathcal C^N_s&=f^{s}_{\q}(\kk)\,f^{s}_{-\q}(\kk'),
\end{split}
\end{equation}
i.e.\ $\mathcal A_s$, $\mathcal B_s$, $\mathcal C_s$ with the second leg $-\kk$ replaced by $\kk'$. The screening part is
Eq.~\eqref{eq:Vrest} with both legs given by the same transition,
\begin{equation}
  P_{\rm rest}(i\nu;\kk,\kk')=\sum_{n\notin\{s\pm\}}\frac{\eta_n\,\big|h_{\q}(\kk)\cdot\phi^n(\q)\big|^2}{i\nu-\eta_n\omega_n},
\end{equation}. The instantaneous terms $V_C^0$ and $V_F$ are left out: they give a
frequency-independent self-energy, which only shifts $\mu$. At $\nu=0$, $P$ is real and non-positive, so $\lambda_Z(0)\ge0$.

\subsection{Matsubara sums}
$V$ in Eq.~\eqref{eq:eliashdisc} depends on the frequencies only through the bosonic frequency transfer
$\nu_d=\omega_n-\omega_m=2\pi Td$, $|d|\leq2N_\omega-1$. We first evaluate $V(i\nu_d;\kk_i,\kk_j)$ at all transfers.
 For $|\nu|$ larger than three
times the largest pole frequency we expand each pole in $1/\nu^2$ and sum the moments of the residues. The poles of $V_{\rm rest}$ are binned on a logarithmic frequency grid.

The kernel is then a convolution in the Matsubara index. A direct application costs $O(N_\omega^2n_{\rm FS}^2)$. We
instead store the kernel at the transfers $\nu_d$, zero-pad it to a length $L\geq6N_\omega$, and apply it by fast
Fourier transform in $O(L\log L\;n_{\rm FS}^2)$. 

The sum is cut off at the band edges.

\subsection{Eigensolver}
We never form the matrix of Eq.~\eqref{eq:eliashdisc}; we only apply it to vectors, and find its six eigenvalues
with the largest real part with a Krylov method. The similarity transformation
$\phi_i(i\omega_n)\to\sqrt{I(\omega_n)/Z(i\omega_n)}\,\phi_i(i\omega_n)$ gives the same eigenvalues and a Hermitian kernel if
\begin{equation}
  V(i\nu;\kk_i,\kk_j)=V(-i\nu;\kk_j,\kk_i)^* .
  \label{eq:hermV}
\end{equation}.
Our interaction satisfies $V(i\nu;\kk_i,\kk_j)=V(i\nu;\kk_j,\kk_i)^*$ instead, and the two agree only if
$\omega_{\q s}=\omega_{-\q s}$. Time-reversal symmetry is broken, so this is not guaranteed. We measure the violation
of Eq.~\eqref{eq:hermV} and use the Lanczos method when it is at round-off and the Arnoldi method otherwise. In the
$\lambda$-jellium model the violation is about $10^{-4}$, and the imaginary parts of the leading eigenvalues are
negligible.

\subsection{Pairs without inversion symmetry}
In rhombohedral graphene the band has no inversion symmetry, so $\kk$ and $-\kk$ on the Fermi surface are not
degenerate. For the pair $(\kk,-\kk)$ let $\xi=E_{\kk}-\mu$, $\xi'=E_{-\kk}-\mu$, $\delta=\xi'-\xi$ and
$\bar\xi=(\xi+\xi')/2$. Moving both electrons by the same $s$ along the normal of the Fermi surface,
\begin{equation}
\begin{split}
  I(\omega)\to\;&\int ds\,\frac{1}{(i\omega-\xi-s)(-i\omega-\xi'-s)}\\
  =\;&\frac1\Omega\Big[\arctan\frac{\xi_{\rm max}-\bar\xi}{\Omega}+\arctan\frac{|\xi_{\rm min}|+\bar\xi}{\Omega}\Big],
\end{split}
\end{equation}
with $\Omega=|\omega|-i\,{\rm sgn}(\omega)\,\delta/2$. It reduces to Eq.~\eqref{eq:Ixi} for $\delta=\bar\xi=0$. For
$|\omega|\ll|\delta|$ the Cooper logarithm is cut off at $|\delta|/2$, which is the pair breaking by the energy
mismatch. $\delta$ varies strongly along the contour, so we average this integral over 16 points on the arc of
each $\kk_i$. The kernel is then complex and not Hermitian, and we use the Arnoldi method. 

\subsection{Critical temperature and gap function}
The interaction does not depend on $T$, so it is built once and Eq.~\eqref{eq:eliashdisc} is solved at a decreasing
list of temperatures. We stop as soon as the leading eigenvalue $\lambda_1(T)$ crosses 1. $T_c$ is obtained by
linear interpolation of $\lambda_1$ in $\ln T$ between the two temperatures that bracket it. If $\lambda_1<1$ at all
temperatures, we extrapolate linearly in $\ln T$ from the two lowest ones. 

Below $T_c$ the linearized equation describes no physical state, and the order of the channels can change with $T$. We therefore solve Eq.~\eqref{eq:eliashdisc} again at $T_c$ and take the gap function from its leading
eigenvector, $\Delta(\kk_i)=\phi_i(i\pi T_c)/m_i$ up to an overall constant, at the lowest positive Matsubara frequency. 

This is how we obtain Fig.\ref{fig:results}(e) and Fig.\ref{fig:r5g}.
\section{Cooper pair angular momentum}
\label{sec:angmom}
The winding number $W$ depends on the Bloch gauge of the lowest HF band: a gauge transformation
$|u_0(\kk)\rangle\to e^{i\theta_\kk}|u_0(\kk)\rangle$ changes the gap to $e^{-i(\theta_\kk+\theta_{-\kk})}\Delta(\kk)$, and a
large gauge transformation around the pocket shifts $W$. The angular momentum of the Cooper pair is gauge-invariant. We write the pair wave function in the plane-wave basis of the parent model and
compute its eigenvalue under the $C_6$ rotation of the crystal.

Let $a^\dagger_\PP$ create an electron in the lowest band of the $\lambda$-jellium model at momentum $\PP$, with
spinor $|\chi_\lambda(\PP)\rangle$. The HF state of the lowest band is
$\hat\gamma^\dagger_0(\kk)=\sum_\G c_0(\kk,\G)\,a^\dagger_{\kk+\G}$, and the Cooper pair is
\begin{equation}
  |F\rangle=\sum_\kk\Delta(\kk)\,\hat\gamma^\dagger_0(\kk)\hat\gamma^\dagger_0(-\kk)|0\rangle
  =\sum_{\PP\PP'}F(\PP,\PP')\,a^\dagger_\PP a^\dagger_{\PP'}|0\rangle ,
\end{equation}
with
\begin{equation}
  F(\PP,\PP')=\sum_\kk\Delta(\kk)\,c_0(\kk,\G)\,c_0(-\kk,\G'),
  \label{eq:Fpair}
\end{equation}where $\PP=\kk+\G,\;\PP'=-\kk+\G'$.
Under the gauge transformation above, $c_0(\kk,\G)\to e^{i\theta_\kk}c_0(\kk,\G)$ and the phases cancel, so $F$ is
gauge invariant.

Let $R$ be the rotation by $\pi/3$. We use the orbital convention $\hat C_6\,a^\dagger_\PP\,\hat C_6^{-1}=a^\dagger_{R\PP}$:
the first component of $|\chi_\lambda(\PP)\rangle$ is real, so the internal degree of freedom carries no angular
momentum. The crystal spontaneously breaks continuous translation, so its rotation centre $\rr_0$, a lattice site,
is wherever the Hartree-Fock solution put it. The rotation about $\rr_0$ is
$\hat C_6(\rr_0)=\hat T_{\rr_0}\hat C_6\hat T_{\rr_0}^{-1}$, which adds a phase to each sector of total momentum
$\mathbf Q=\PP+\PP'$:
\begin{equation}
  \big[\hat C_6(\rr_0)F\big](R\PP,R\PP')=e^{i(\mathbf Q-R\mathbf Q)\cdot\rr_0}\,F(\PP,\PP') .
\end{equation}
The pair angular momentum $L$ is defined by
\begin{equation}
  \hat C_6(\rr_0)|F\rangle=e^{-iL\pi/3}|F\rangle ,\qquad L\ ({\rm mod}\ 6),
  \label{eq:Ldef}
\end{equation}
and we compute it from the overlap
\begin{equation}
  \frac{\langle F|\hat C_6(\rr_0)|F\rangle}{\langle F|F\rangle}
  =\frac{\sum_{\PP\PP'}F^*(R\PP,R\PP')\,e^{i(\mathbf Q-R\mathbf Q)\cdot\rr_0}\,F(\PP,\PP')}{\sum_{\PP\PP'}|F(\PP,\PP')|^2}.
\end{equation}
We find $\rr_0$ by maximizing the modulus of this overlap over the unit cell. 

\subsection{Relation to the winding number}
The gap function in the main text is evaluated in the pocket-smooth gauge, defined by parallel transport of the
Bloch states from the band extremum at $\Gamma$. Rotations commute with this transport, so a rotated Bloch state
equals the rotated state at $\Gamma$ times a fixed phase,
$\hat C_6(\rr_0)|\psi_0(\kk)\rangle=e^{-iL_\Gamma\pi/3}|\psi_0(R\kk)\rangle$, where $L_\Gamma$ is the $C_6$ eigenvalue of the
band-0 state at $\Gamma$ in the same orbital convention. The pair contains two electrons, and a gap with winding
$W$ contributes $e^{-iW\pi/3}$, so
\begin{equation}
  L=W+2L_\Gamma\quad({\rm mod}\ 6).
\end{equation}
We compute $L_\Gamma$ independently from the rotation eigenvalue of the band-0 state at $\Gamma$ about the same
$\rr_0$, and this relation holds for every eigenvector we have computed. Across the band inversion $W$ jumps by 2
and $L_\Gamma$ by $-1$, so $L$ is unchanged.

\section{Rhombohedral graphene}
\subsection{Hamiltonian}
We use the model of rhombohedral pentalayer graphene of Ref.~\cite{Feng2026SDC} for a single spin- and
valley-polarized flavor. The Hamiltonian is
\begin{equation}
  \hat H=\sum_{\PP}a^\dagger_{\PP}\,h(\PP)\,a_{\PP}
  +\frac{1}{2\mathcal A}\sum_{\q}V(\q)\,{:}\hat\rho_{\q}\hat\rho_{-\q}{:},\qquad
  V(\q)=\frac{2\pi e^2}{\epsilon_r\,q}\tanh(qd),
\end{equation}
with $d=25$\,nm, $\epsilon_r=15$, and the normal ordering taken relative to charge neutrality. The momentum $\PP$ is
measured from the graphene valley $K$, and $a_{\PP}=(a_{\PP A_1},a_{\PP B_1},\dots,a_{\PP A_5},a_{\PP B_5})$ runs over the
sublattices $A_\ell,B_\ell$ of the layers $\ell=1,\dots,5$. With $\pi=P_x+iP_y$ and $v_i=\frac{\sqrt3}{2}a\,t_i$
($a=0.246$\,nm), the nonzero matrix elements of the continuum Hamiltonian $h(\PP)$ are
\begin{eqnarray}
  &&h_{A_\ell B_\ell}=v_0\,\pi^*,\qquad
  h_{B_\ell A_{\ell+1}}=t_1,\qquad
  h_{A_\ell A_{\ell+1}}=h_{B_\ell B_{\ell+1}}=-v_4\,\pi^*,\nonumber\\
  &&h_{A_\ell B_{\ell+1}}=-v_3\,\pi,\qquad
  h_{A_\ell B_{\ell+2}}=\frac{t_2}{2},\nonumber\\
  &&h_{A_\ell A_\ell}=-u_D(\ell-3)-\Delta'\,\delta_{\ell 1},\qquad
  h_{B_\ell B_\ell}=-u_D(\ell-3)-\Delta'\,\delta_{\ell 5},
\end{eqnarray}
together with their Hermitian conjugates. This is the small-$\PP$ expansion of the lattice model of
Ref.~\cite{Feng2026SDC} around $K$, with $(t_0,t_1,t_2,t_3,t_4)=(3100,380,-21,290,141)$\,meV, $\Delta'=10.5$\,meV for
the dimer-site crystal field, $u_D$ the displacement-field potential, and the moir\'e potential switched off. The
on-site term agrees with theirs up to a constant and the sign convention of $u_D$. In the Hartree-Fock calculation we keep the $N_b=10$ lowest
conduction bands of $h$.

\paragraph{Inversion within the valley.} Inversion and $C_{2z}$ both exchange the valleys, so nothing in a
valley-polarized state requires $E(\PP)=E(-\PP)$. An intravalley inversion would be a unitary $U$ with
$U\,h(-\PP)\,U^\dagger=h(\PP)$. The intralayer ($t_0$) and dimer ($t_1$) terms fix it, up to a sign, to the
diagonal $U_{A_\ell}=(-1)^{\ell-1}$, $U_{B_\ell}=(-1)^{\ell}$. Under this $U$, the $t_4$ terms (odd in $\PP$, between
orbitals of opposite sign) and the on-site terms are invariant. Two terms are not:
\begin{itemize}
\item the trigonal warping $t_3$: $h_{A_\ell B_{\ell+1}}=-v_3\pi$ is odd in $\PP$ but connects orbitals of the same sign;
\item the next-nearest-layer hopping $t_2$: $h_{A_\ell B_{\ell+2}}=t_2/2$ is even in $\PP$ but connects orbitals of
  opposite sign.
\end{itemize}
The symmetric part $\frac12[T(\PP)+U\,T(-\PP)\,U^\dagger]$ of either term vanishes identically. To restore the
inversion symmetry within the valley we therefore set
\begin{equation}
  t_2=t_3=0,
\end{equation}
keeping all other terms and parameters. Then $U\,h(-\PP)\,U^\dagger=h(\PP)$ holds exactly, and since $V(\q)$ is even
in $\q$, the Hartree-Fock functional, the collective modes and the pairing kernel inherit the symmetry, unless the
self-consistent state breaks it spontaneously, which we check. In particular $E(\kk)=E(-\kk)$ on the Fermi
surface, and $C_3$ combined with $\kk\to-\kk$ acts as $C_6$ on the Fermi contour.

\subsection{Cooper pair angular momentum}
In rhombohedral graphene each plane wave also carries a layer and sublattice index $f$, and the crystal is only
$C_3$ symmetric, so $L$ is defined mod 3. The continuum Hamiltonian of one valley satisfies
$h(R\PP)=D\,h(\PP)\,D^\dagger$ for $R$ the rotation by $2\pi/3$, with $D$ diagonal in $f$: $D_{A_\ell}=\omega^{\ell-1}$
and $D_{B_\ell}=\omega^{\ell}$ on layer $\ell$, $\omega=e^{2\pi i/3}$. A rotation therefore acts as
$\hat C_3\,a^\dagger_{\PP f}\,\hat C_3^{-1}=D_f\,a^\dagger_{R\PP f}$, and the pair picks up $D_fD_{f'}$ in addition to the
phase above. The overall phase of $D$ is a convention. We fix $D_{A_1}=1$: the conduction band edge lies on the $A_1$
orbital, so, as in the $\lambda$-jellium model, the orbital on which the band lives at its edge carries no
internal angular momentum. The triangular crystal has three inequivalent $C_3$ centres per unit cell, the site
$\rr_0$ and the two triangle centres $\rr_0+\mathbf t$ with $(1-R)\mathbf t$ a lattice vector. For a pair of zero
total crystal momentum, $\mathbf Q$ is a reciprocal lattice vector and
$(\mathbf Q-R\mathbf Q)\cdot\mathbf t=-(R\mathbf Q)\cdot(1-R)\mathbf t\in2\pi\mathbb Z$, so $L$ is the same about all three centres.